# Achieving translational symmetry in trapped cold ion rings

Hao-Kun Li[1*], Erik Urban[2*], Crystal Noel[2], Alexander Chuang[2], Yang Xia[1], Anthony Ransford[2], Boerge Hemmerling[2], Yuan Wang[1,3], Tongcang Li[1], Hartmut Häffner[2,3], Xiang Zhang[1,3]

[1]NSF Nanoscale Science and Engineering Center, 3112 Etcheverry Hall, University of California, Berkeley, California 94720, USA

[2]Department of Physics, University of California, Berkeley, California 94720, USA

[3]Materials Sciences Division, Lawrence Berkeley National Laboratory, 1 Cyclotron Road, Berkeley, CA 94720, USA

Correspondence and requests for materials should be addressed to H.H. (hhaeffner@berkeley.edu) and X.Z. (xiang@berkeley.edu)

*These authors contributed equally to this work.

**Spontaneous symmetry breaking is a universal concept throughout science. For instance, the Landau-Ginzburg paradigm of translational symmetry breaking underlies the classification of nearly all quantum phases of matter and explains the emergence of crystals, insulators, and superconductors[1]. Usually, the consequences of translational invariance are studied in large systems to suppress edge effects which cause undesired symmetry breaking[2]. While this approach works for investigating global properties, studies of local observables and their correlations require access and control of the individual constituents. Periodic boundary conditions, on the other hand, could allow for translational symmetry in small systems where single particle control is achievable. Here, we crystallize up to fifteen $^{40}$Ca$^+$ ions in a microscopic ring with inherent periodic boundary conditions. We show the ring's translational symmetry is preserved at millikelvin temperatures by delocalizing the Doppler laser cooled ions. This establishes an upper bound for undesired symmetry breaking at a**

**level where quantum control becomes feasible. These findings pave the way towards studying quantum many-body physics with translational symmetry at the single particle level in a variety of disciplines from simulation of Hawking radiation[3] to exploration of quantum phase transitions[4].**

Engineered quantum systems with periodic boundary conditions have led to observation of many interesting physical phenomena and enabled important applications over the past decades[5-7]. However, quantum control of the individual constituents remains very challenging in such systems. Atomic trapped ions, on the other hand, offer a large degree of control down to the single-particle level[8-10]. In light of this, the unique properties of trapped ion rings open up novel opportunities to study diverse topics. For instance, the ring geometry confines topological defects while still allowing them to move freely. This property extends the lifetime of the quasiparticles relevant for studies of the Kibble-Zureck mechanism[11,12] and dynamics of kink solitons[13,14]. The ring topology also allows one to study the Aharanov-Bohm effect in the quantum regime[15] and symmetry breaking with indistinguishable particles[16,17]. A rotating ion ring lends itself to studies of the acoustic analogue of Hawking radiation by introducing a controlled phonon dispersion[3]. Additionally, because each ion experiences the same potential, the system could be advantageous in quantum computation[18], frequency metrology[19], and allow one to study quantum frictions[20] and quantum phase transitions[4] in a homogeneous setting.

Ion rings have been previously implemented by concatenating conventional linear ion trap designs into a circle[21-24]. In these experiments, the ion-electrode distances have been much smaller than the ion ring diameters, making the ring crystals sensitive to complex stray electric fields from nearby surfaces. Although the stray fields do not significantly affect high-energy ion rings[23], they substantially disrupt the translational symmetry at low temperatures where many interesting quantum phenomena can be observed. Hence it is desirable to reduce the ion-ring diameter while keeping the ions far away from the electrodes, such that stray fields from imperfections vary on length scales larger than the ring diameter. This strategy avoids local distortions of the ion ring crystal. Here, using a novel trap design[25,26], we crystallize $^{40}$Ca$^+$ ions in a 90 μm diameter ring with 390 μm ion-electrode distance, a 60-fold increase of the ratio between ion-electrode distance and the ring diameter. The new design allows compensation of the unavoidable stray fields with primarily homogeneous fields in order to achieve a high degree of translational symmetry.

Our ring trap consists of three concentric circular electrodes surrounded by eight static-voltage compensation electrodes, as shown in Fig. 1a. Applying a radio-frequency (rf) voltage to the innermost and outermost circular electrodes with all other electrodes held at a dc voltage generates a time-averaged circular potential minimum above the trap surface[26] (Fig. 1b). The trap is fabricated from Boron doped Silicon anodically bonded on borofloat glass ($SiO_2$), as shown in Fig. 2a. The electrodes are formed using photolithography followed by deep dry etching of silicon (Fig. 2b). The deep electrode trenches (Fig. 2c) ensure that stray fields from bound charges in the glass are well shielded. Hydrofluoric acid (HF) etching of the glass underneath the trenches is performed to increase the distance between the electrodes through the glass surface. This process prevents surface breakdown when a high voltage is applied between neighboring electrodes[27]. In the interest of leaving the potential above the trap undisturbed, electrical vias are created on the backside by HF etching of the glass substrate, followed by Gold deposition and a lift-off process.

The ring trap is operated by applying a 220 V amplitude, $2\pi \times 5.81$ MHz signal to the rf electrodes. A flux of neutral Calcium atoms generated from a heated atom oven travels parallel to the trap surface in the *x*-direction, as shown in Fig. 2d. The neutral Calcium is then ionized inside the trapping region through a two-photon process at 422 nm and 375 nm. A red detuned 397 nm laser beam incident from the *y*-direction cools the ions while an 866 nm laser beam repumps the ions out of the dark D-state[28]. The fluorescence of the ions at 397 nm is collected with a custom objective[29] and imaged upon an electron multiplying charge coupled device (EMCCD) camera.

When captured by the trapping potential, the ions crystallize into a ring because of their mutual Coulomb repulsion, as shown in Fig. 3. With no compensating fields applied, the ions are typically pinned to one side of the ring by stray electric fields. The strength of the stray fields in the *x-y* plane is measured to be ~ 3 V/m by recording the compensating field required to reposition the ion crystal to be first *x* then *y* axis symmetric. The measured radial trapping frequency is $2\pi \times 390$ kHz (Fig. 3a). The trapping potential is able to hold up to fifteen ions in a ring before the ion crystal forms a zig-zag configuration when pinned.

The presence of external electric fields in the trapping plane creates an asymmetry in the ring potential, resulting in a finite tangential trapping frequency (Fig. 3a). We gauge the asymmetry by measuring the tangential trapping frequency of the ion crystal. In the measurement, we apply a sinusoidal voltage to one compensation electrode and observe the excitation of the collective tangential mode with the EMCCD camera[30]. Figures 3e and 3f present the observed dependence of the tangential trapping frequency on the total external dipole field strength $E_y$ and the ion number *N*, respectively. As the ion number becomes larger, the increased Coulomb repulsion resulting from the reduced ion-ion spacing enforces a more uniform charge distribution in the ring, as shown in Figs. 3a and 3c. For such a homogeneous charge distribution, an external electric field exerts a smaller restoring torque when the ion crystals deviate from the equilibrium position. Therefore, we expect the tangential trapping frequencies to decrease with increasing ion number and better compensation of the stray field.

The potential energy of the ion crystals can be modeled by considering only a homogeneous external field and the coulomb repulsion of the ions confined to a ring. This results in a potential energy of the form $V = -\sum_i \frac{1}{2} E_y ed \cos\theta_i + \sum_{i<j} e^2 / (4\pi\varepsilon_0 d \sin|\frac{\theta_i - \theta_j}{2}|)$, where $d$ denotes the ring diameter, $\theta_i$ the angular position of the $i$ th ion, $e$ the elementary charge and $\varepsilon_0$ the vacuum permittivity. We calculate the frequencies of the collective tangential motion by expanding the potential energy of the ion crystals to quadratic order of the ion displacements relative to equilibrium. The results are presented in Fig. 3e and 3f. The calculation agrees with the experimental results over the full extent of the measurement without free fitting parameters. This result confirms that homogeneous electric fields are the dominant symmetry breaking mechanism at the energy scales of our measurement.

In order to quantify the scale at which the symmetry is broken, we use the magnitude of the perturbations of the potential that causes the localization of the ion ring, defined as the rotational energy barrier. To obtain the rotational energy barrier, we numerically vary the position of the final ion in the chain and use the model confirmed above to solve for the lowest potential energy configuration of the remaining ions, as described by Figs. 4a-c. The calculated potential energy of a ten-ion crystal versus the final ion position is plotted in Fig. 4d. The two minimum energy locations correspond to the original equilibrium configuration of the crystal and the energy peak represents the rotational energy barrier $V_B$. For ten ions, we observe localized ring crystals with in-plane electric fields larger than $(2.0 \pm 0.1)$ V/m (Fig. 3e). At this electric field, the rotational energy barrier is calculated to be $V_B/k_B = (10 \pm 4)$ mK, where $k_B$ is the Boltzmann constant.

By decreasing the external field or increasing the ion number, the rotational energy barrier can be reduced to the point where the ion ring delocalizes, as shown in Fig. 3d. For a ten-ion crystal, delocalization occurs at $E_y = -(1.9 \pm 0.1)$ V/m (Fig. 4e), corresponding to a rotational energy barrier of $V_B/k_B = (6 \pm 3)$ mK. We find that the rotational energy barrier at which the ion ring starts to delocalize is independent of the number of ions, as shown in Fig. 4f. In addition, we determine the tangential temperature of the ions to be ~ 3 mK by measuring the Doppler-broadened width of the 729 nm $4^2S_{1/2}$–$3^2D_{5/2}$ transition[27]. The proximity of the ion temperature to $V_B/k_B$ suggests that delocalization occurs when the thermal energy of the ions is large enough to overcome the rotational energy barriers.

Ultimate control of the ion ring requires translational symmetry of its rotational ground state in the picokelvin regime. In the future, we look to explore what mechanism could break the symmetry at energy scales below the Doppler limit studied here. So far we have found that the main symmetry breaking mechanism is homogeneous fields. From our model, we estimate that with our achieved control of ~ 0.1 V/m, homogeneous fields will not affect the symmetry of a ten-ion ring even at the $10^{-12}$ K level. Approaching these small energies, we expect that higher order multipole fields will become relevant. To explore these effects, further cooling of the rotational degree-of-freedom is required.

**Methods**

Deep dry etching of the silicon was performed with a STS deep reactive ion etching system, using SPR-220 photoresist. Before etching of the vias, the glass substrate was thinned down from 175 μm to 70 μm using HF, and a metal etching mask consisting of 60 nm Chromium and 200 nm Gold was patterned by photolithography. Before assembling in the ultra-high vacuum chamber, the trap was coated with thin a thin layer of Silver using electron-beam evaporation from the top and was bonded to a printed circuit board with solder paste. The SEM image of the trap cross-section was recorded with an LEO 1550 microscope.

**Acknowledgements**

We thank SUSS MicroTec and Applied Microengineering Ltd for wafer bonding services. The authors thank Norman Yao for helpful discussions. This project is supported by WM Keck Foundation. E.U. acknowledges support by the NSF Graduate Research Fellowship under Grant No. 1106400.

**Author Contributions**

X.Z., H.H., and T.L. initiated the project. H.-K.L., Y.X., T.L., and Y.W. fabricated the ring trap. C.N., E.U., H.-K.L., A.R., and A.C. setup the experiment. E.U. and H.-K.L. performed the measurement. A.C. and H.-K.L. performed numerical calculation. H.-K.L., E.U., C.N., B.H., H.H., and X.Z. analyzed the results and prepared the manuscript. H.H. and X.Z guided the research. All authors contributed to the discussions.

**Author Information**

The authors declare no competing financial interests.

Correspondence and requests for materials should be addressed to hhaeffner@berkeley.edu and xiang@berkeley.edu.

**Figure 1**

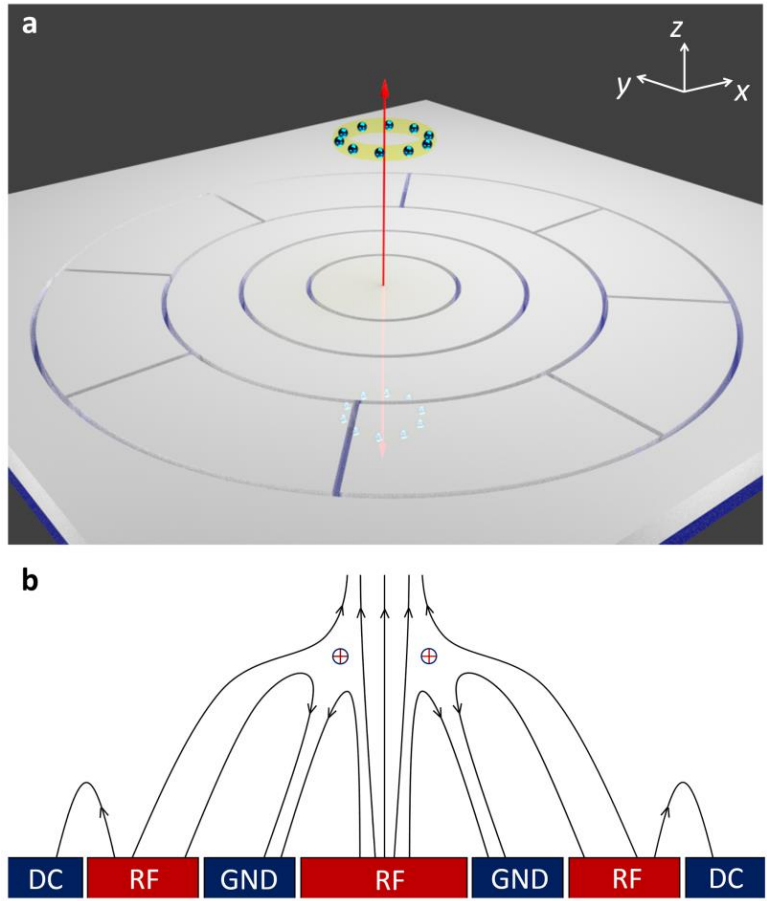

**Figure 1 | Schematics of the surface-electrode ring trap. a**, The trap consists of three circular electrodes and eight static voltage electrodes in a plane. The outer radius of the three circular electrodes are 125 μm, 600 μm, and 1100 μm, respectively. The gap between the innermost and the next circular electrode is 15 μm, and the gaps between other electrodes are 25 μm. The whole electrode pattern possesses a diameter of 6 mm, outside of which is ground. Applying rf voltages to the innermost and outermost circular electrodes generates a time-averaged ring shaped electrical potential which enables the formation of the ring ion crystals. **b**, Cross-sectional electric fields at an instant when the applied rf potential is positive relative to ground.

**Figure 2**

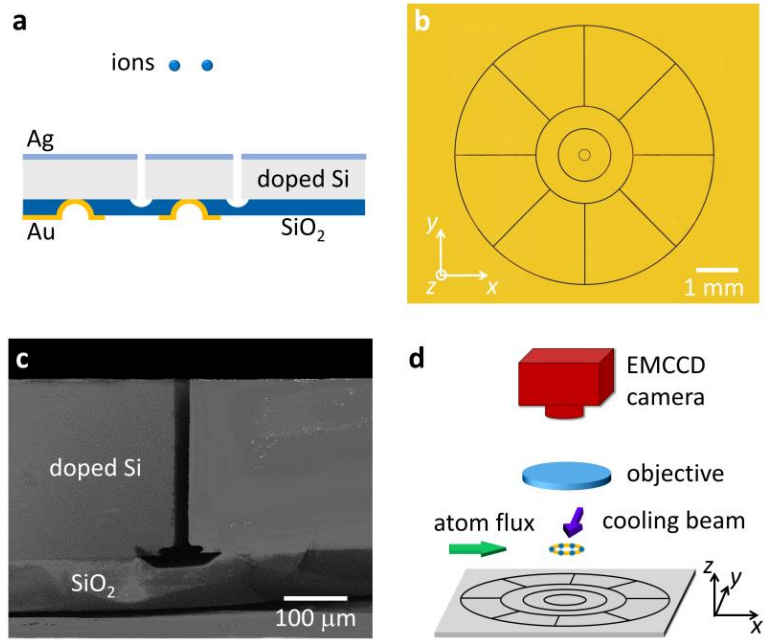

**Figure 2 | Trap fabrication and experiment setup. a**, Cross-sectional schematics of the fabricated trap. The doped silicon electrodes coated with Silver are mechanically supported by an insulating SiO$_2$ substrate. The resistivity of the doped silicon is < 0.005 ohm·cm. The thickness of the Ag, Si, and SiO$_2$ layers are 100 nm, 250 μm, and 70 μm, respectively. Electrical vias through the SiO$_2$ layer are coated with a 800 nm layer of Gold. **b**, Stitched optical image of the trap electrodes illuminated with yellow light. **c**, Scanning electron microscope image of the trap cross-section. The deep trench in silicon separates two electrodes. The cut beneath the trench prevents surface flashover when high voltage is applied between the electrodes. **d**, Experimental setup for observing the ion crystals. The neutral atom flux generated by a heated atom oven travels parallel to the trap surface in the *x*-direction and then ionized with a two-photon process in the trapping regime. The cooling beam parallel to the trap surface comes from the *y*-direction.

**Figure 3**

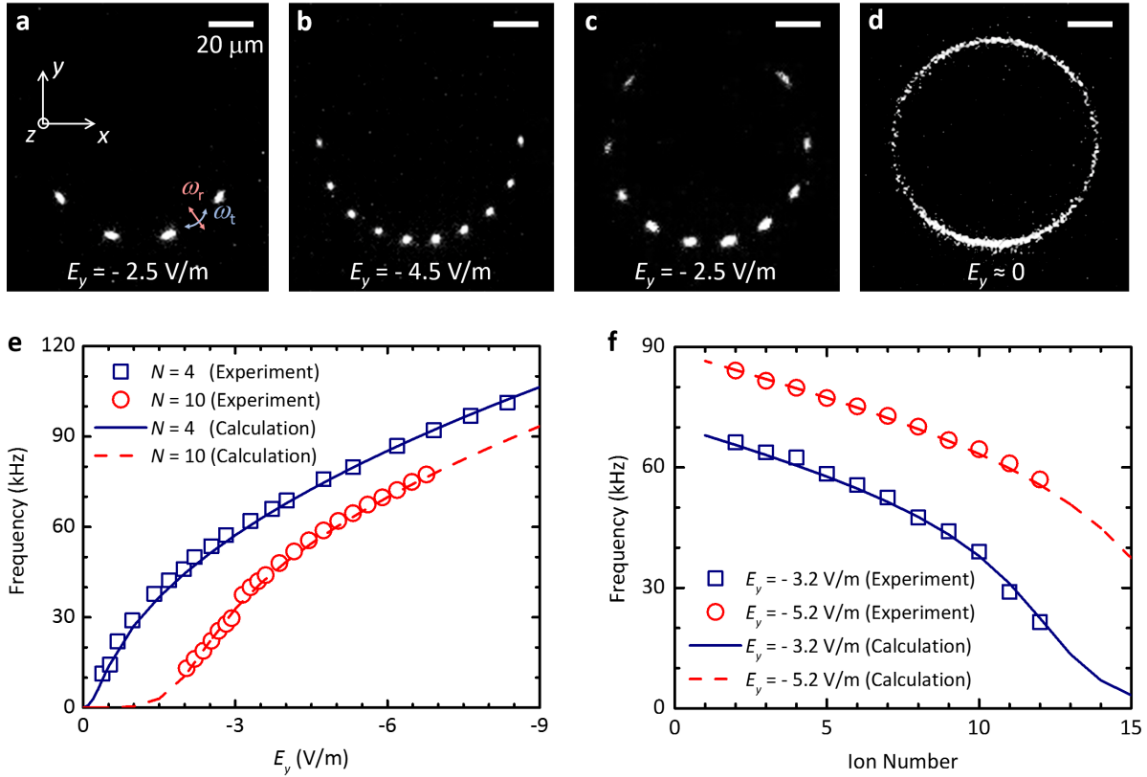

**Figure 3 | Fluorescence images of ring ion crystals and measurement of tangential trapping frequencies. a**-**c**, Images of ion crystals composed of four ions (**a**) and ten ions (**b** and **c**). In **a**, the red and blue arrows depict radial ($\omega_r$) and tangential ($\omega_t$) trapping directions. **d**, Image of a delocalized ten-ion ring when the total external dipole field is close to zero (exposure time 200 ms). In **a-d**, the scale bars are 20 μm. The flourescence inhomogeneity of the images is caused by the size of the Gaussian cooling beam of ~ 70 μm full width at half maximum. **e**, **f**, Dependence of the tangential trapping frequency on the total external dipole field strength $E_y$ and ion number $N$, respectively. Error bars are smaller than the sizes of the data points. The lines correspond to the calculated collective tangential frequencies using the electric potential model considering only a homogeneous electric field and the Coulomb repulsion of the ions confined to a ring. The

agreement between the calculated results and the experimental data confirms that the homogenous electric fields are the dominant symmetry breaking mechanism.

**Figure 4**

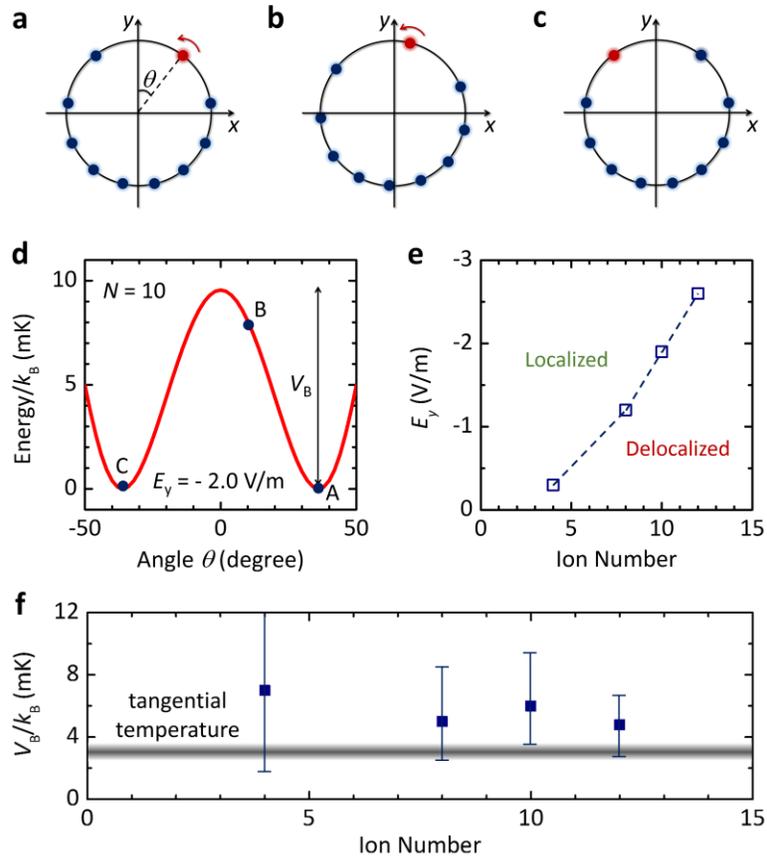

**Figure 4 | Analysis of the rotational energy barrier. a-c**, Lowest potential energy configurations of the remaining nine ions varying with the position of the final ion (marked in red) in a ten-ion crystal. The original equilibrium configuration is shown in **a**. When the final ion moves from right to left, the crystal recovers the original equilibrium configuration, as shown in **c**. **d**, Calculated potential energy of ten ions as a function of the final ion position angle $\theta$, where the difference between the minimum and the maximum is the rotational energy barrier $V_B$. Points A,

B, and C correspond to the equilibrium configurations shown in **a**, **b**, and **c**, respectively. In **a-d**, $E_y = -2.0$ V/m. **e, f**, Total external electric field strength (**e**) and corresponding rotational energy barrier $V_B/k_B$ (**f**) at which the ion crystals are observed to delocalize as a function of the ion number. The errors of the electric fields in **e** are $\pm 0.1$ V/m. The gray line in **f** denotes the measured tangential temperature ~ 3 mK of the ions, which is close to $V_B/k_B$, and suggests that the delocalization occurs when the thermal energy of the ions is large enough to overcome the rotational energy barrier.